\documentclass[aps, pra, reprint, longbibliography,
superscriptaddress]{revtex4-1}

\usepackage{hyperref}
\usepackage{graphicx}
\usepackage[utf8]{inputenc}
\usepackage{xcolor}
\usepackage{amsmath, amssymb}
\usepackage{physics}
\usepackage{ulem}
\usepackage{epstopdf}

\begin{document}

\preprint{APS/123-QED}

\title{Quantum antidipolar systems in two-dimensional geometries
}

\author{J. S\'anchez-Baena}
\email{juan.sanchez.baena@upc.edu}
\affiliation{Departament de F\'isica, Universitat Polit\`ecnica de Catalunya, Campus Nord B4-B5, 08034 Barcelona, Spain}

\author{J. Boronat}
\email{jordi.boronat@upc.edu}
\affiliation{Departament de F\'isica, Universitat Polit\`ecnica de Catalunya, Campus Nord B4-B5, 08034 Barcelona, Spain}

\date{\today}

\begin{abstract}
Particles with magnetic moment
can be polarized, and rapidly rotated, employing a magnetic field 
such that the dipolar interaction among them changes sign, becoming 
\textit{antidipolar} and thus isotropically attractive in a plane. Polar 
molecules can also be manipulated using microwave dressing fields to invert the 
sign of the dipole-dipole interaction. In this work, we study
a two-dimensional system of antidipolar particles by calculating its equation 
of state and structural properties. 
The system behaves as a liquid even for scattering lengths significantly 
greater than the dipolar length. For large enough densities, the system 
transitions to a solid with one particle per lattice site via a first-order 
phase transition at a significantly smaller density than its dipolar 
counterpart. Moreover, motivated by the recent realization of a strongly 
axially trapped, bilayer geometry [Science \textbf{384}, 546-551 (2024)], we 
study the properties of the bilayer liquid phase as the inter-layer distance is 
tuned, and provide the range of parameters where one layer can influence the 
properties of the other.

\end{abstract}

\maketitle

\section{\label{sec:introduction}Introduction}

Ultracold atoms constitute an experimentally feasible and highly controllable 
platform to study a plethora of physical phenomena. Among them, dipolar systems 
have enjoyed a great interest in the 
recent years due to their remarkable combination of long-range character and 
anisotropy. This mix gives rise to the celebrated dipolar 
supersolids~\cite{Modugno:PRL:2019,Pfau:PRX:2019,norcia21:nature}, which combine 
diagonal and off-diagonal long range orders. Most theoretical and experimental 
works on the field have focused on 
the study of the bare atomic dipole-dipole interaction (DDI), although the high 
controllability achieved in dipolar systems offers the possibility to tune the 
strength and even the sign of the DDI. This can be achieved in the context of 
magnetic atoms by polarizing and rapidly rotating 
them~\cite{giovanazzi02:prl,tang18:prl,kirby23:SciPostPhysCore}, achieving a 
reversal in the sign of the DDI for a tilting angle of $\phi = 90^o$ with 
respect to the rotation axis. It is also possible to achieve fully antidipolar 
interactions in the context of the recently realized polar 
molecules~\cite{bigagli24:nature,shi25:arxiv,deng2025:natcomm,tijs2025:prxq}, 
which are able to attain dipolar strengths several orders of magnitude larger 
than magnetic atoms. In this context, changing the ellipticity of one of the 
applied microwave fields (which are necessary to avoid excessive losses) allows 
to tune the strength and sign of the DDI. The alteration of the sign of the DDI 
leads to major changes in supersolid formation, especially under anisotropic 
traps~\cite{halder22:prr,mukherjee25:pra}, to the formation of pancake shaped 
droplets~\cite{zhang26:nature,langen25:prl,ciardi25:prl}, as well as to the 
stabilization of dark solitons in 3D~\cite{nath08:prl} and bright solitons in 
2D~\cite{pedri05:prl}. On the other hand, in the context of mixtures 
composed by dipolar and non-dipolar atoms, the 
reversal of the sign of the DDI has been shown to change the order of the 
supersolid transition~\cite{kirby23:SciPostPhysCore}.

\begin{figure}[t]
\centering
\includegraphics[width=\linewidth]{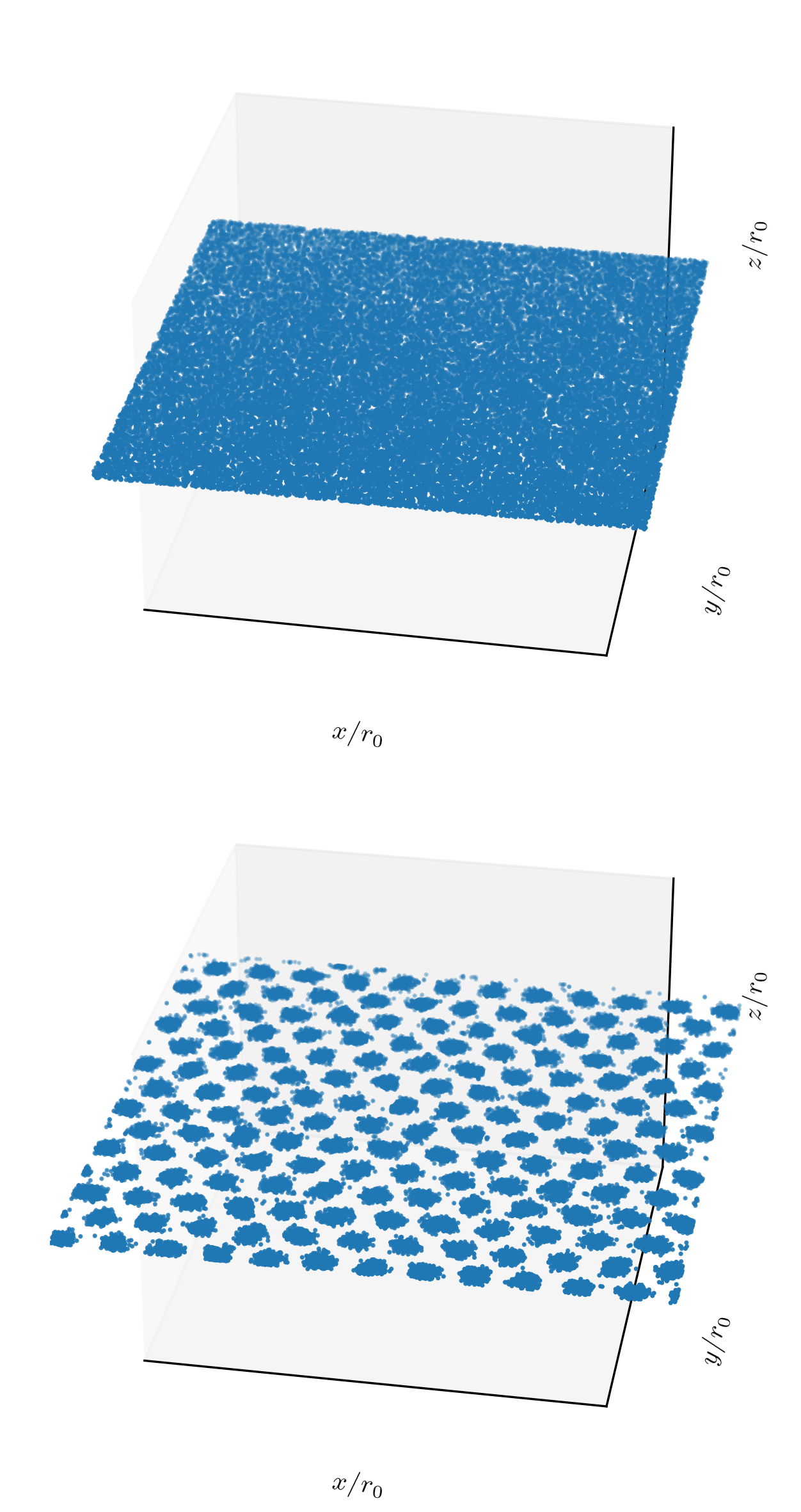}
\caption{Simulation snapshots of the liquid (top) and solid (bottom) phases in 
the antidipolar system. The strength of the repulsive core is set to 
$C_{12}/(E_0 r_0^{12}) = 0.02073$ (see Eq.~\ref{potential}) while the density in 
each case is $n r_0^2 = 0.17$ (top) and $n r_0^2 = 2.5$ (bottom).}
\label{fig0}
\end{figure}

The change in the sign of the DDI becomes especially relevant when other sources 
of anisotropy, like the confinement, are introduced. In two dimensions (2D), a 
polarized dipolar system can be globally or partially repulsive depending on the 
tilting angle of the dipoles with respect to the normal direction to the plane. 
Instead, for antidipolar particles, the interaction becomes purely attractive. 
This increased attraction may facilitate the phase transition to a crystal 
phase (see Fig.~\ref{fig0}), where one atom sits in each of the sites of a 
perfect triangular lattice. This transition is known to take place in 2D systems 
of regular dipoles at exceedingly high densities~\cite{macia14}, while an 
antidipolar crystal of polar molecules in a finite system has already been 
obtained via quantum Monte Carlo calculations~\cite{ciardi25:prl}. Moreover, the 
strong antidipolar interaction in molecules has been found to induce the 
formation of 2D structures without the need of any external 
confinement~\cite{langen25:prl}. However, the full characterization of the 
liquid-solid transition in the thermodynamic limit is still lacking.

In recent years, many efforts have been directed towards the realization of 
ultracold atomic systems in quasi-two dimensional (quasi-2D) 
configurations~\cite{sunami2022:prl,Yu2024:science,Guo2024:science,
liao2026:arxiv}, including dipolar 
systems~\cite{he25:scadv,zhen2025:arxiv,he2025:arxiv,he2026:arxiv}. In the 
context of two dimensional physics, a significant interest has been directed to 
the experimental realization of bilayer 
geometries~\cite{beregi24:avs,lidu24:science,rydow25:natcomm,chang25:arxiv,
lin26:arxiv}, which have been extensively studied from a theoretical 
viewpoint~\cite{pikovski10:prl,matveeva11:eurphys,matveeva13:prl, 
safavi-Naini13:njp,camacho16:pra,vanzyl16:pra,mazloom17:prb,guijarro2021:scirep, 
guijarro22:prl,renklioglu24:pra,tiene24:pra,zhang25:prb}.
If the inter-particle interactions are of long-range character, this opens the 
possibility to 
the presence of inter-layer effects, where the presence of one layer affects the 
properties of the other. In the context of dipolar interactions, the long-range 
property of the inter-layer interactions can couple the motion of the center of 
mass of two separated clouds~\cite{matveeva11:eurphys}, influence the polaronic 
properties of an impurity~\cite{matveeva13:prl}, induce inter-layer 
drag~\cite{renklioglu24:pra} or induce the formation of a liquid 
phase~\cite{guijarro22:prl}. Bilayer systems of dipolar particles are 
interesting even beyond the context of ultracold atoms, since they serve as a 
model for indirect excitonic systems in 
semiconductors~\cite{hubert19:prx,choksy21:prb}. With these precedents, we aim 
to investigate to which extent the antidipolar interaction
can give rise to inter-layer effects that modify the equilibrium properties of one of the layers due to the presence of the other.

In this work, we characterize a two-dimensional system of antidipolar particles 
in single 
and bilayer geometries. We study the equilibrium properties of the system in its 
ground state, mainly the energy and its spatial structure. With this 
information, we study the liquid-solid phase transition for different 
interaction parameters in the single layer case. In the bilayer geometry, we 
focus in the liquid phase and study the inter-layer effects in the energy per 
particle of the system and the condensate and superfluid fractions of the other 
layer. We contextualize our calculations in the framework of potential 
experiments with magnetic atoms taking the setup of Ref.~\cite{lidu24:science} 
as a reference, while calculations featuring the recently developed models for 
the interactions between polar 
molecules~\cite{deng2025:natcomm,tijs2025:prxq} will be the subject of a future 
work.

\section{\label{sec:system}System and Hamiltonian}

\subsection{\label{sec:system1}Single layer}

We study a two-dimensional system of $N$ magnetic atoms under periodic boundary conditions described by the Hamiltonian
\begin{align}
 \hat{H} = \sum_{i=1}^N -\frac{\hbar^2 \nabla_i^2}{2m} + \sum_{i=1}^N \sum_{j>i}^N V({\bf r}_i - {\bf r}_j) \ ,
 \end{align}
where the inter-particle potential is given by
\begin{align}
 V({\bf r}) = \frac{C_{12}}{\abs{ {\bf r} }^{12}} - \frac{C_{\rm dd}}{4 \pi \abs{ {\bf r} }^3} \ ,
 \label{potential}
\end{align}
where ${\bf r} = x \hat{i} + y \hat{j}$ is a two-dimensional vector. 
The first term ($\frac{C_{12}}{\abs{ {\bf r} }^{12}}$) models a generic short 
range repulsion while the last term ($- \frac{C_{\rm dd}}{4 \pi \abs{ {\bf r} 
}^3}$) corresponds to the antidipolar interaction. The strength of the
former can be adjusted to the scattering length in potential experimental 
realizations, while the latter can be achieved in systems of magnetic atoms by 
rapidly rotating magnetic 
dipoles~\cite{giovanazzi02:prl,tang18:prl,kirby23:SciPostPhysCore} or in systems 
with polar molecules under single and double microwave shielding by setting 
appropriate values of the Rabi frequency and the detuning of the microwave 
fields~\cite{deng2025:natcomm,tijs2025:prxq}. In the case of magnetic dipoles, 
$C_{\rm dd}$ is then given by $C_{\rm dd} = \mu_0 \mu^2 / 2$ (notice the factor 
of 2 dividing the magnetic moments, which arises from the time average of the 
rotation of the dipoles). We express our results in dipolar units, with the 
characteristic length and energy given by $r_0 = \frac{m C_{\rm dd}}{4 \pi 
\hbar^2}$ and $E_0 = \frac{\hbar^2}{m r_0^2}$. As an example, for $^{164}$Dy 
atoms $r_0 = 196.5$ $a_0$ while for $^{166}$Er atoms $r_0 = 98.25$ $a_0$, 
where $a_0$ is the Bohr radius.

\begin{figure}[t]
\centering
\includegraphics[width=\linewidth]{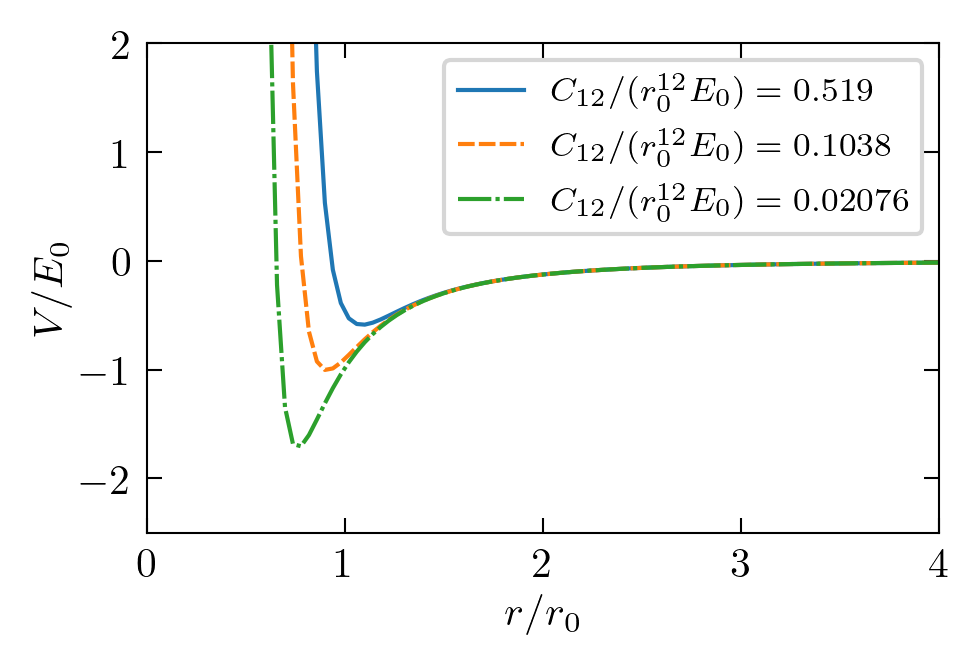}
\caption{Antidipolar potential (see Eq.~\ref{potential}) for different values of $C_{12}$.}
\label{fig1}
\end{figure}

In experimental realizations, two-dimensional physics is explored in a
quasi-2D setting, where the axial trapping $\omega$ is strong enough for 
``strict 2D'' physics to arise (like the BKT transition), but weak enough such 
that $a_{\rm 3D} < l$, where $l$ is the harmonic oscillator length and $a_{\rm 
3D}$ is the 3D scattering 
length~\cite{he25:scadv,zhen2025:arxiv,he2025:arxiv,he2026:arxiv}. These 
conditions imply that the microscopic collisions between particles can be 
treated as 3D events~\cite{pilati05:pra}, and thus the relevant interaction 
parameter in experiments is the 3D scattering length. In the following, we 
explore the energetic and structural properties of the system for different 
values of the short-range repulsive strength $C_{12}$. We consider the values 
$C_{12}/(E_0 r_0^{12}) = 0.0207, 0.1038, 0.519$ which yield 3D scattering 
lengths of $a_{\rm 3D}/r_0 = 0.4, 0.496, 0.605$, respectively. If we compare 
these values with the dipole length $a_{\rm dd} = m C_{\rm dd}/(12 \pi \hbar^2) 
= r_0/3$, we can see that, in all cases, $a_{\rm 3D} > a_{\rm dd}$. We show in 
Fig.~\ref{fig1} an illustration of the inter-particle potential for the 
different values of $C_{12}$ considered.

\subsection{\label{sec:system2}Bilayer}

For the bilayer geometry, our system is composed by $N = N_1 + N_2$ particles, 
all with the same mass, in 
layers 1 and 2, with $N_1 = N_2 = N/2$ particles in each layer. The layers are 
separated a distance $h$ in the $z$-axis. The Hamiltonian is given by
\begin{align}
 \hat{H} &= \sum_{i=1}^N -\frac{\hbar^2 \nabla_i^2}{2m} + \sum_{i=1}^{N_1}\sum_{j>i}^{N_1} V^{11}({\bf r}_i - {\bf r}_j) \nonumber \\
 & + \sum_{i=1}^{N_2} \sum_{j>i}^{N_2} V^{22}({\bf r}_i - {\bf r}_j) + \sum_{i=1}^{N_1} \sum_{j=1}^{N_2} V^{12}({\bf r}_i - {\bf r}_j) \ ,
\end{align}
where $V^{11} = V^{22}$ and $V^{12}$ are the intra and inter-layer potentials, 
respectively, which are given by
\begin{align}
 V^{11}({\bf r}) &= \frac{C_{12}}{\abs{ {\bf r} }^{12}} - \frac{C_{\rm dd}}{4 \pi \abs{ {\bf r} }^3} \\
 V^{12}({\bf r}) &= \frac{C_{12}}{ \left( \abs{\bf r}^2 + h^2 \right)^{6}} - \frac{C_{\rm dd}}{4 \pi \left( \abs{\bf r}^2 + h^2 \right)^{3/2}} \nonumber \\
 &\times \left[ 1 - 3 \frac{h^2}{ \left( \abs{\bf r}^2 + h^2 \right) } \right] \ .
\end{align}
In the calculations for the bilayer geometry, we fix the value of $C_{12}$ to 
$C_{12}/(E_0 r_0^{12}) = 0.1038$. 
Notice that, while the intra-layer potential is always negative at long 
distances, the inter-layer potential $V^{12}({\bf r})$ can be either positive 
or negative depending on the anisotropic term $1-3 \cos^2\theta=1 - 3 
\frac{h^2}{ \left( \abs{\bf r}^2 + h^2 \right) }$. This interaction features a 
repulsive cone if $\abs{\bf r} < \sqrt{2} h$ and otherwise it is attractive.

\subsection{\label{sec:method} Quantum Monte Carlo method}

We perform our study using the diffusion Monte Carlo method 
(DMC)~\cite{chin90}. 
This method numerically solves the imaginary-time evolution equation
\begin{align}
 &\psi_T(\mathbf{R}) \psi(\mathbf{R}, \tau + \Delta \tau) = \nonumber \\
 &\int d\mathbf{R'} G(\mathbf{R},\mathbf{R'},\Delta \tau) \frac{\psi_T(\mathbf{R})}{\psi_T(\mathbf{R'})} \psi_T(\mathbf{R'}) \psi(\mathbf{R'}, \tau) \ , \label{imag_ev}
\end{align}
where $\tau = i t/\hbar$ is the imaginary time, $G(\mathbf{R},\mathbf{R'},\Delta 
\tau) = \bra{\mathbf{R}} \exp \left( - \hat{H} \Delta \tau \right) 
\ket{\mathbf{R'}}$ is the imaginary time Green's function and 
$\psi_T(\mathbf{R})$ is the trial wave function, an input to the method. This 
trial wave function helps to reduce the variance of the estimations if chosen 
appropriately. In practice, the numerical implementation of Eq.~\ref{imag_ev} 
works as follows: one first obtains a statistical representation of the density 
associated to a given trial wave function, $\rho_T(\mathbf{R}) = 
\abs{\psi_T(\mathbf{R})}^2$. This representation consists of a set of points in 
coordinate space named \textit{walkers} (i.e. $\{ \vec{R} \}_i = \{ 
\vec{r}_1\text{, ..., } \vec{r}_N\}_i$ where $i$ is the walker index). A set of 
transformations are then applied to the walkers, such that, at the end of the 
process, they statistically represent the probability distribution 
$\psi_0(\mathbf{R}) \psi_T(\mathbf{R})$, with $\psi_0(\mathbf{R})$ the ground 
state wave function of the system. Observables are estimated as
\begin{align}
 \langle \hat{O} \rangle = \frac{\int d\mathbf{R} \text{ } \psi_T(\mathbf{R}) \hat{O} \psi_0(\mathbf{R}) }{ \int d\mathbf{R} \text{ } \psi_T(\mathbf{R}) \psi_0(\mathbf{R})} \ .
\end{align}
Notice that any observable that commutes with the Hamiltonian can be computed 
exactly (up to statistical uncertainty) regardless of the choice of the trial 
wave function. 

We employ different trial wave functions for the single and bilayer cases
\begin{align}
\psi_T^{\rm single}(\mathbf{R}) &= F_1({\bf R})\prod_{i=1}^N \prod_{i<j}^N f_{11}\left(r_{i j}\right) \label{trial_Psi_single}
 \\
\psi_T^{\rm bilayer}(\mathbf{R})&=\left( \prod_{i=1}^{N_1} \prod_{i<j}^{N_1} f_{11}\left(r_{i j}\right) \right) \left( \prod_{i=1}^{N_2} \prod_{i<j}^{N_2} f_{22}\left(r_{i j}\right) \right) \nonumber \\
&\times \left( \prod_{i=1}^{N_1} \prod_{j=1}^{N_2}  f_{12}\left(r_{ij}\right) \right)
\label{trial_Psi_bilayer}
\end{align}
where $r_{ij} = \abs{\mathbf{r}_{ij}}$. In the single layer case, for our 
parameters of choice, the system can be either in the liquid or solid phase. 
The difference in translational symmetry is captured by the one-body factor 
$F_1({\bf R})$, which is different for the solid and liquid phases, i.e.,
\begin{align}
 F_1({\bf R}) &=
 \begin{cases}
  \sum_{i=1}^N \exp(-\alpha ({\bf r}_i - {\bf r}^c_i)^2) & \text{ solid phase }
  \\
  1 & \text{ liquid phase }
 \end{cases}
\end{align}
where $\alpha$ is a variational parameter (to be optimized via the variational 
Monte Carlo method for 
every density) and ${\bf r}^c_i$ are the 2D position vectors of a triangular 
lattice. The factors $f_{11} = f_{22}$ in 
Eqs.~\ref{trial_Psi_single}~\ref{trial_Psi_bilayer} are the intra-layer Jastrow 
factors and $f_{12}$ is the inter-layer one. These are obtained by solving the 
two-body problem, which differs in the single and bilayer cases. For the
parameters of interest in this work, the two-body ground state wave functions
$\Psi_{\rm 2body}(r_{ij})$ correspond to
weakly bound states. If, for a given simulation box of length $L$, the global
maximum of $\Psi_{\rm 2body}(r_{ij})$ sits at a position $r_{\rm max} < L/2$, the
Jastrow factor is taken as
\begin{align}
  f(r_{ij}) =
 \begin{cases}
 \Psi_{\rm 2body}(r_{ij}) & \text{ if } r_{ij} < r_{\rm max}
 \\
 \Psi_{\rm 2body}(r_{\rm max})  & \text{ if } r_{ij} \ge r_{\rm max} \ .
 \end{cases}
\end{align}
Otherwise, if $r_{\rm max} > L/2$, we compute the Jastrow factor as
\begin{align}
  f(r_{ij}) =
 \begin{cases}
 \Psi_{\rm 2body}(r_{ij}) & \text{ if } r_{ij} < r^{*}
 \\
 C_2 \exp \left[ C_1 \left( -\frac{1}{r_{ij}} +\frac{1}{L - r_{ij}} \right) 
\right] & \text{ if } r^{*} < r_{ij} < L/2
 \\
 1 & \text{ if } r_{ij} \ge L/2
 \end{cases}
\end{align}
where $C_1$ and $C_2$ are found by imposing the continuity of $f(r_{ij})$ and 
its first derivative at $r_{ij} = r^{*}$ and we set $r^{*} = 0.9 (L/2)$ as 
optimal value. The analytical form chosen in the range $r_{ij} \in [r^{*}, 
L/2]$ guarantees the necessary continuity of the Jastrow factor and its 
first derivative for all values of $r_{ij}$.

For the single layer simulations, we set the particle number to $N = 100$ in the 
liquid phase
and $N = 168$ in the solid phase. The simulation box, which features periodic 
boundary conditions, is squared in the liquid phase ($L_y = L_x = L$) and 
slightly asymetric ($L_y = 1.01 L_x = L$) in the solid phase due to the geometry 
of the triangular lattice. In the bilayer simulations we only focus in the 
liquid phase and set $N_1 = N_2 = 50$ and a squared simulation box. To 
accurately compute the energy of the system, we need to account for finite size 
effects due to the finiteness of the simulation box. This is especially relevant 
for the antidipolar interaction since it is quasi long-range in two dimensions. 
To do so, we follow the same prescription as Ref.~\cite{guijarro22:prl}, i.e. we 
use the relation between the interaction energy and the pair distribution 
function $g(r)$, which, in general, is given by $E_{\rm inter}/N = (n/2) 
\int_0^{\infty} V(r) g(r) d{\bf r}$. We then compute this integral for $r \in 
[L/2, \infty]$ setting the intra-layer and inter-layer pair distribution 
functions to unity, i.e., $g_{11}(r) = g_{22}(r) = g_{12}(r) = 1$. This yields 
a correction to be added to the energies obtained with DMC. The energy 
corrections are
\begin{align}
 E_{\rm corr}/N =
 \begin{cases}
  -\frac{2 \pi n}{L} & \text{single layer}
  \\
  -\pi n \left( \frac{2}{L} + \frac{\left( L/2 \right)^2}{ \left( \left( L/2 \right)^2 + h^2 \right)^{3/2} } \right) & \text{bilayer}
 \end{cases}
 \label{e_corr}
\end{align}
where, for the case of the bilayer, $n$ is the 2D density of a single layer. One can check that the expressions in Eq.~\ref{e_corr} match those of Ref.~\cite{guijarro22:prl} with the exception of a change in sign, which comes from the inverted sign of the antidipolar interaction.

\section{\label{sec:results}Results}


\subsection{\label{sec:single_layer}Single layer}

\begin{figure}[t]
\centering
\includegraphics[width=\linewidth]{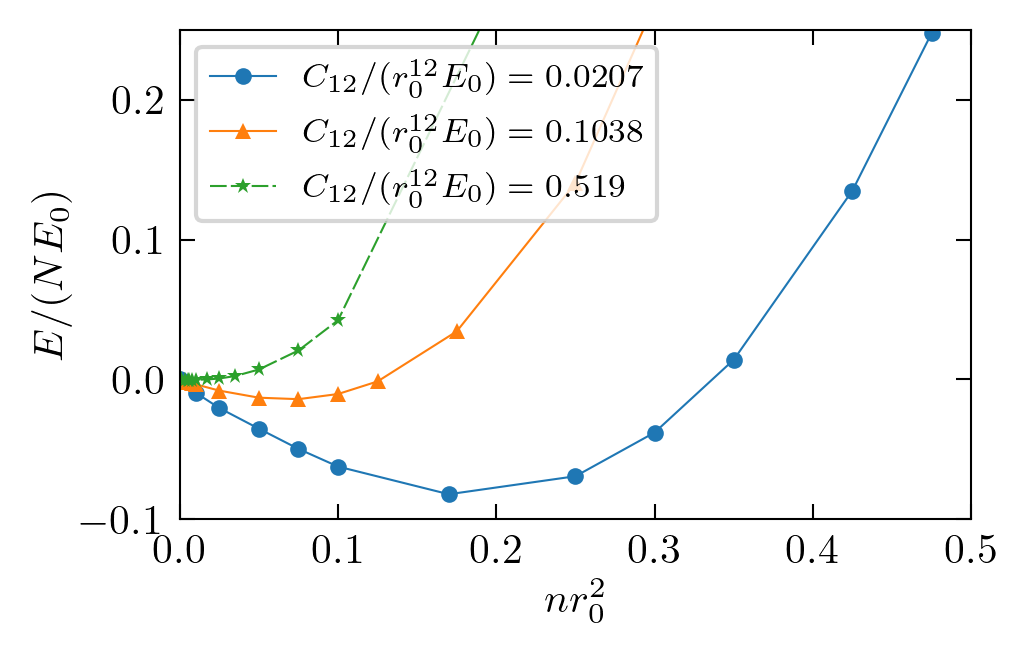}
\caption{Energy per particle as a function of the 2D density for the single 
layer, 2D antidipolar liquid for different values of the interaction parameter 
$C_{12}$ (see Eq.~\ref{potential}).}
\label{fig2}
\end{figure}

We first characterize the energetic properties of the two-dimensional 
antidipolar system. In particular, 
the equation of state (EOS) provides insightful information on its self-bound 
character. 
In Fig.~\ref{fig2}, we report our DMC results for the EOS for different values 
of $C_{12}$. The energy per particle shows 
a clear minimum at a finite, non-zero density where the condition of mechanical 
equilibrium (zero pressure) is satisfied.  As $C_{12}$ increases, the 
interaction becomes more repulsive and consequently, the equilibrium density of 
the liquid decreases. For the parameters considered, the equilibrium density of 
the liquid lies in the range between $n r_0^2 = 0.01$ and $n r_0^2 = 0.17$.

We can relate these equilibrium 2D densities with the corresponding 3D peak 
densities that would
arise in a potential experimental realization to have an idea of their order of 
magnitude. If one considers a quasi-2D geometry strongly trapped along the 
$z$-axis, the wave function can be reasonably splitted as $\psi_{\rm 3D}({\bf 
r}) = \psi_{\rm 2D}({\bf r}) \left( \frac{1}{\pi l^2} \right)^{1/4} 
e^{-\frac{1}{2}(z/l)^2}$.
Under these conditions, for the density, we have $n_{\rm 3D}({\bf r}) = n 
\sqrt{\frac{1}{\pi l^2}} e^{-(z/l)^2}$, which implies that given a 2D density 
$n$ the 3D peak density is given by $n_{\rm 3D, peak} = n \sqrt{\frac{1}{\pi 
l^2}}$. For a range of 2D densities, between $n r_0^2 = 0.01$ and $n r_0^2 = 
0.17$, we obtain a range of peak 3D densities of $n_{\rm 3D, peak} \in [2.5 
\times 10^{15}, 4 \times 10^{16}]$ cm$^{-3}$ for the axial trap used in the 
experiment of Ref.~\cite{lidu24:science} (where $T > T_{\rm BKT}$) and 
considering $^{164}$Dy atoms. This means that the peak density range is above the threshold $n_{\rm 3D, peak}
> 10^{15}$ cm$^{-3}$,
 meaning that three-body losses become significant over tens of
miliseconds~\cite{zhang21:pra,he25:prr}. The maximum 3D gas parameter is
$x_{\rm 3D, max.} \sim 10^{-3}$ for the trap of Ref.~\cite{lidu24:science},
which is close to the dilute regime.
Also, for this trap, $l \simeq 2 r_0$, which implies $\hbar \omega > E/N$ around the
equilibrium density of the liquid, validating the quasi-2D condition.

For large enough densities, the system experiments a first order phase 
transition 
to a crystal phase with coexistence densities $n_l^*$ and $n_s^*$ for the liquid 
and the solid, respectively. At these densities, mechanical and chemical 
equilibrium takes place, i.e. $P_l = P_s$ and $\mu_l = \mu_s$, where $P$ is the 
pressure and $\mu$ is the chemical potential. As it is common in two-dimensional 
systems, these coexistence densities are remarkably close, $n_l^* \simeq 
n_s^*$~\cite{Whitlock88:pra}. To illustrate this, we compute these densities for
$C_{12}/(E_0 r_0^{12}) = 0.0207$. In order to obtain them, we calculate $E/N$ 
for several values of the 2D density for the liquid ($n_l$) and solid ($n_s$) phases, and
fit the data to the phenomenological functionals:
\begin{align}
 \varepsilon_l(n_l) &= \varepsilon_l^{\rm eq} + a_l (n_l-n_l^{\rm eq})^{b_l} \label{fit_l} \\
 \varepsilon_s(n) &= a_s + b_s n_s^{c_s} \label{fit_s}
\end{align}
where $\varepsilon_l^{\rm eq}$ and $n_l^{\rm eq}$ are the equilibrium energy per particle and density of the liquid, respectively. The parameters $a_l$, $b_l$, $a_a$, $b_s$ and $c_s$ are fitting parameters. From these fits, we can obtain the pressure and chemical potential of the liquid and the crystal phases, respectively
\begin{align}
 P_l &= n_l^2 \eval{ \pdv{\varepsilon_l}{n} }_{N} = n_l^2 b_l a_l (n_l - n_l^{\rm eq})^{b_l - 1} \\
 \mu_l &= \varepsilon_l + P_l/n_l \\
 P_s &= n_s^2 \eval{ \pdv{\varepsilon_s}{n} }_{N} = c_s b_s n_s^{c_s + 1} \\
 \mu_s &= \varepsilon_l + P_s/n_s \ .
\end{align}
We numerically find the values of the densities of the liquid and the crystal at 
which $P_l \simeq P_s$ and $\mu_l \simeq \mu_s$ and take these as our 
estimations for the coexistence densities $n_l^*$ and $n_s^*$.

\begin{figure}[t]
\centering
\includegraphics[width=\linewidth]{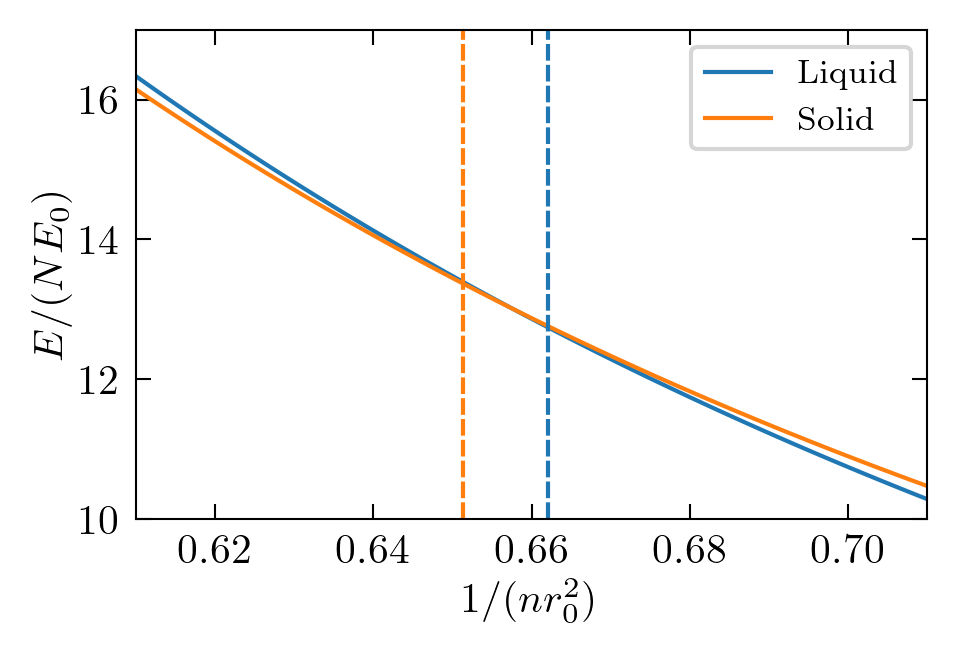}
\caption{Fits of the energy per particle of the solid and crystal phases (see 
Eqs.~\ref{fit_l}~\ref{fit_s}) as a function of the inverse 2D density $1/n$ for 
different values of $C_{12}/(E_0 r_0^{12}) = 0.02076$.}
\label{fig3}
\end{figure}

We show in Fig.~\ref{fig3} the energy per particle $E/N$ as a function of the 
inverse of the density $1/n$ in a region around the energy crossing point for 
$C_{12}/(E_0 r_0^{12}) = 0.0207$, indicating the value of the coexistence 
densities with a dashed line. Due to their proximity, $n_l^*$ and $n_s^*$ lay 
remarkably close to the density of the energy crossing between the two phases, 
$n_{\rm cross}$. Therefore, we only report this value in Table~\ref{table1} for 
all the interactions considered. Increasing $C_{12}$ makes the crossing 
densities decrease, and thus the lattice constant of 
the crystal, $r_{ij}^{\rm crystal}$, increase. In the crystal phase, the 
positions of the particles are centered around the classical positions of a 2D 
triangular lattice. It is expected that these positions are close to the 
points where the potential is minimum, $r_{ij}^{\rm min}$. Looking at 
Fig.~\ref{fig1} this seems to be the case, as this minimum shifts to slightly 
larger values as $C_{12}$ increases. We show in Table~\ref{table1} a comparison 
between $r_{ij}^{\rm crystal}$ and $r_{ij}^{\rm min}$ at the crossing density 
for all the values of $C_{12}$. As we can see from the results, the 
inter-particle distances of a classical 2D triangular lattice are close (around 
a $\sim 10 \%$ apart only) from the minimum inter-particle potential positions.
\begin{figure*}[t]
\centering
\includegraphics[width=0.8\linewidth]{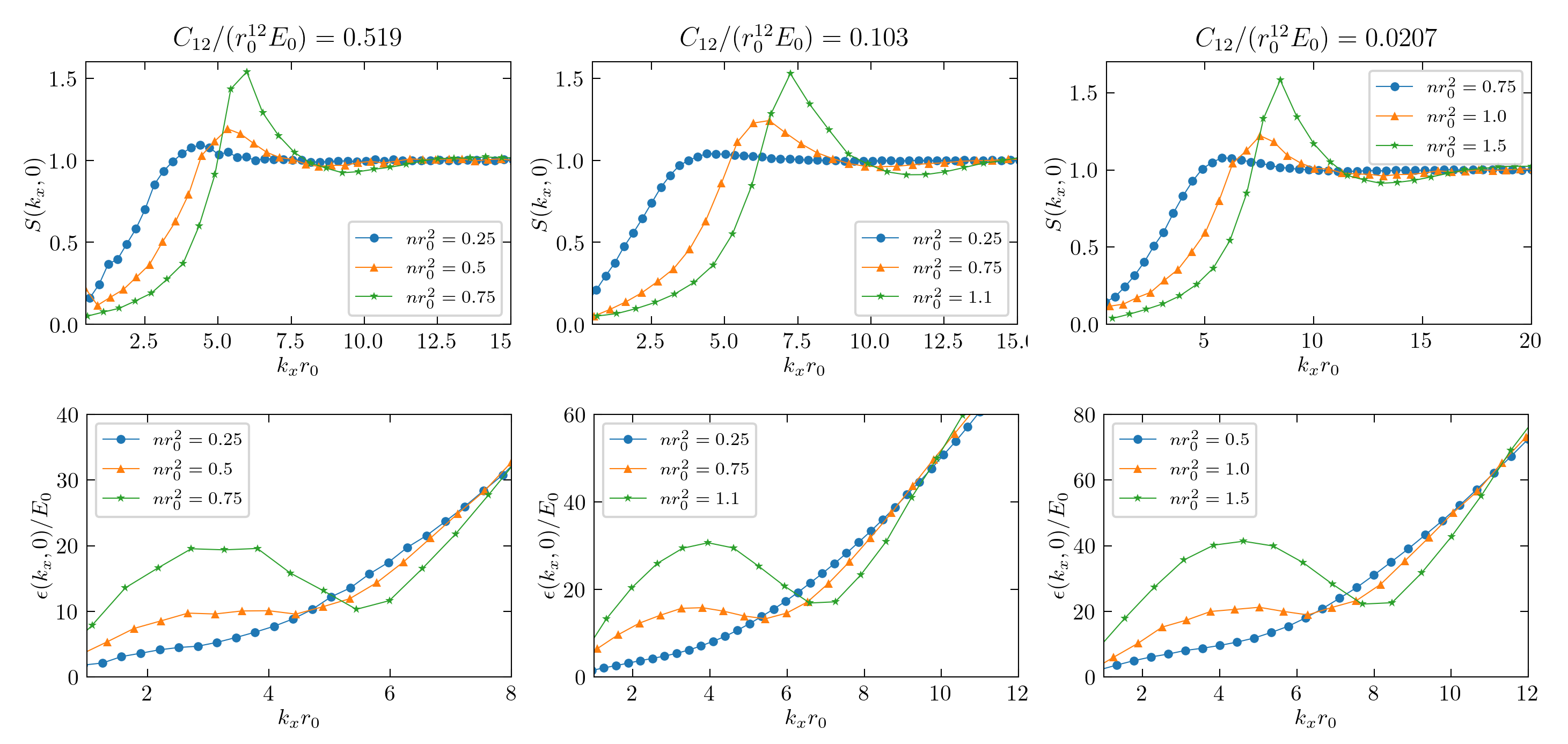}
\caption{Static structure factor (top) and excitation spectrum $\varepsilon(k_x,0)$ (bottom) of the liquid phase for different densities and $C_{12}/(E_0 r_0^{12}) = 0.519 \text{ (left)}, 0.103 \text{ (center)}, 0.0207 \text{ (right)}$.}
\label{fig8}
\end{figure*}
\begin{center}
\begin{tabular}{ | c | c | c | c | c | }
  \hline $C_{12}$ & $n_{\rm cross} r_0^2$ & $r_{ij}^{\rm crystal}/r_0$ & $r_{ij}^{\rm min.}/r_0$  \\ \hline
 $0.0207$ & $1.52$ & $0.86$ & $0.75$ \\ \hline
 $0.103$ & $1.1$ & $1.02$ & $0.91$ \\ \hline
 $0.519$ & $0.8$ & $1.20$ & $1.08$ \\ \hline
\end{tabular}
 \label{table1}
\end{center}
From the results we see that the crossing densities are of the order $n_{\rm cross} r_0^2 \sim 1$, which yields an average inter-particle separation of $d \sim r_0$. Remarkably, this crossing density is about a hundred times smaller than the one found in a system of purely repulsive, conventional 2D dipoles polarized perpendicularly to the plane~\cite{macia14}.
This crossing density implies a 3D peak density of $n_{\rm 3D, peak} \simeq 2.5 
\times 10^{17}$ cm$^{-3}$
for the trap used in the $^{164}$Dy experiment of Ref.~\cite{lidu24:science}.

In strictly 2D and zero temperature the condensate fraction has a finite value.
We can estimate this important quantity by looking at the 
long-range behavior of the one-body density matrix, which is defined as
\begin{equation}
\rho_1({\bf r}) = N {\int d{\bf r}_2 \cdots d{\bf r}_N
\Psi^*({\bf r}_1 + {\bf r}, {\bf r}_2, \cdots, {\bf r}_N)
\Psi({\bf r}_1, {\bf r}_2, \cdots, {\bf r}_N)
\over
\int d{\bf r}_1 \cdots d{\bf r}_N
\mid\! \Psi({\bf r}_1, \cdots, {\bf r}_N) \!\mid^2
} \ .
\label{rho1}
\end{equation}
The condensate fraction is then obtained as $n_0/n=\rho_1(|{\bf 
r}|\to\infty)/n$~\cite{macia11:pra}. At the equilibrium density of the liquid, we find $n_0/n = 0.700 \pm 0.005 \text{, } 0.767 \pm 0.005$ and $0.912 \pm 0.005$, while at the liquid-solid crossing density, we obtain $n_0/n = 0.045 \pm 0.005 \text{, } 0.043 \pm 0.005$ and $0.040 \pm 0.005$ for $C_{12}/(E_0 r_0^{12}) = 0.0207\text{, } 0.103$ and $0.519$, respectively. This means that the condensate is
finite but almost 
entirely depleted at the transition density to the solid phase.
The liquid-solid transition is also signaled by the emergence of a peak 
in the static structure factor of the liquid as its density approaches the 
coexistence density of the phase transition. This is illustrated in 
Fig.~\ref{fig8}, where we report $S(k_{\xi}) = \frac{1}{N} \langle \left( 
\sum_{i=1}^N e^{i k_i {\xi}} \right) \left( \sum_{i=1}^N e^{-i k_i {\xi}} 
\right) \rangle$ where $\xi = x,y$ for different values of the density and 
$C_{12}/(E_0 r_0^{12}) = 0.0207, 0.103, 0.519$. Employing the Feynman 
approximation, $\varepsilon({\bf k}) = \frac{\hbar^2 k^2}{2 m S({\bf k})}$, we 
can estimate a rigorous upper bound to the excitation spectrum of the system at 
low momenta from the static structure factor. The emergence of a peak in 
$S({\bf k})$ induces a roton minimum in the excitation spectrum, which is a 
precursor of spatial ordering~\cite{chomaz2018:natphys}.

\begin{figure}[t]
\centering
\includegraphics[width=\linewidth]{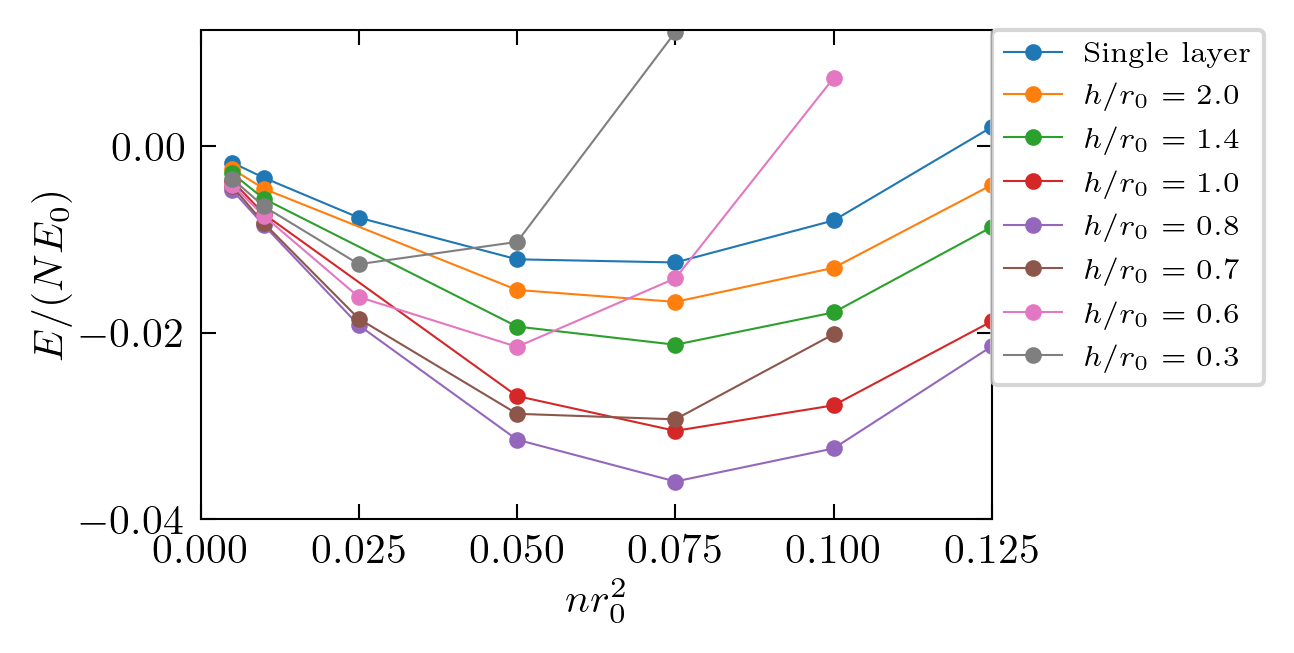}
\caption{Energy per particle as a function of the 2D density for the 2D antidipolar bilayer liquid and for different values of the inter-layer separation $h$. The parameter $C_{12}$ is set to $C_{12}/(E_0 r_0^{12}) = 0.103$ (see Eq.~\ref{potential}).}
\label{fig4}
\end{figure}

\subsection{\label{sec:bilayer}Bilayer}

We now move on to the study of a bilayer system of 2D antidipoles in the 
thermodynamic limit. 
We restrict ourselves to the case where both layers, labelled as ``1'' and ``2'' 
are equaly populated, which guarantees mechanical equilibrium due to the 
intra-layer interactions being the same for both layers. Also, we set 
$C_{12}/(E_0 r_0^{12}) = 0.103$ for all the calculations in this section.

\begin{figure}[t]
\centering
\includegraphics[width=\linewidth]{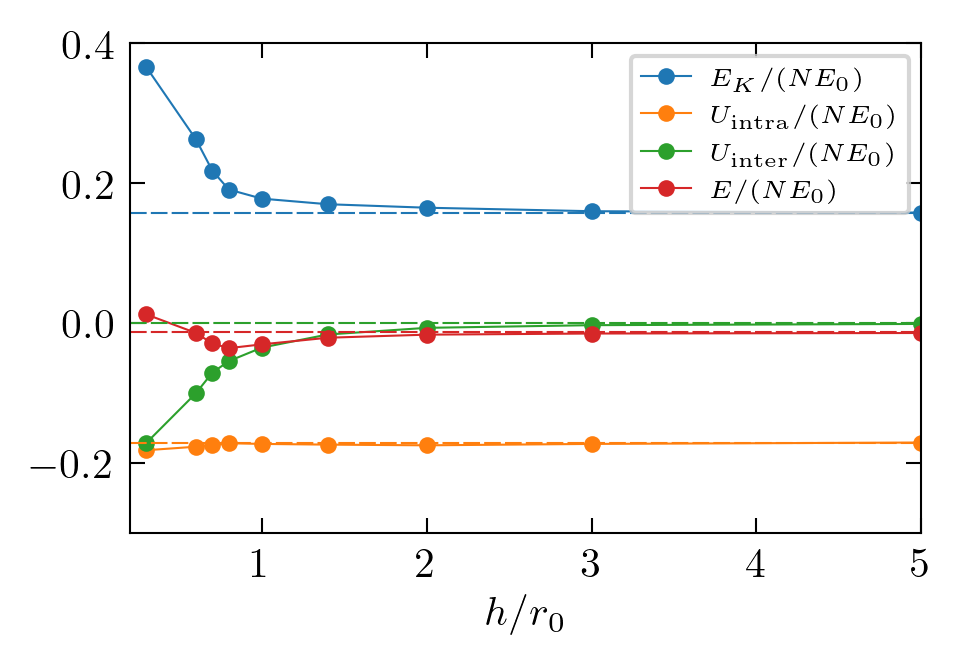}
\caption{Contributions to the bilayer energy per particle for different values of $h$ for $n r_0^2 = 0.075$ and $C_{12}/(E_0 r_0^{12}) = 0.103$. The dashed lines correspond to the single layer case.}
\label{fig4b}
\end{figure}

\begin{figure*}[t]
\centering
\includegraphics[width=\linewidth]{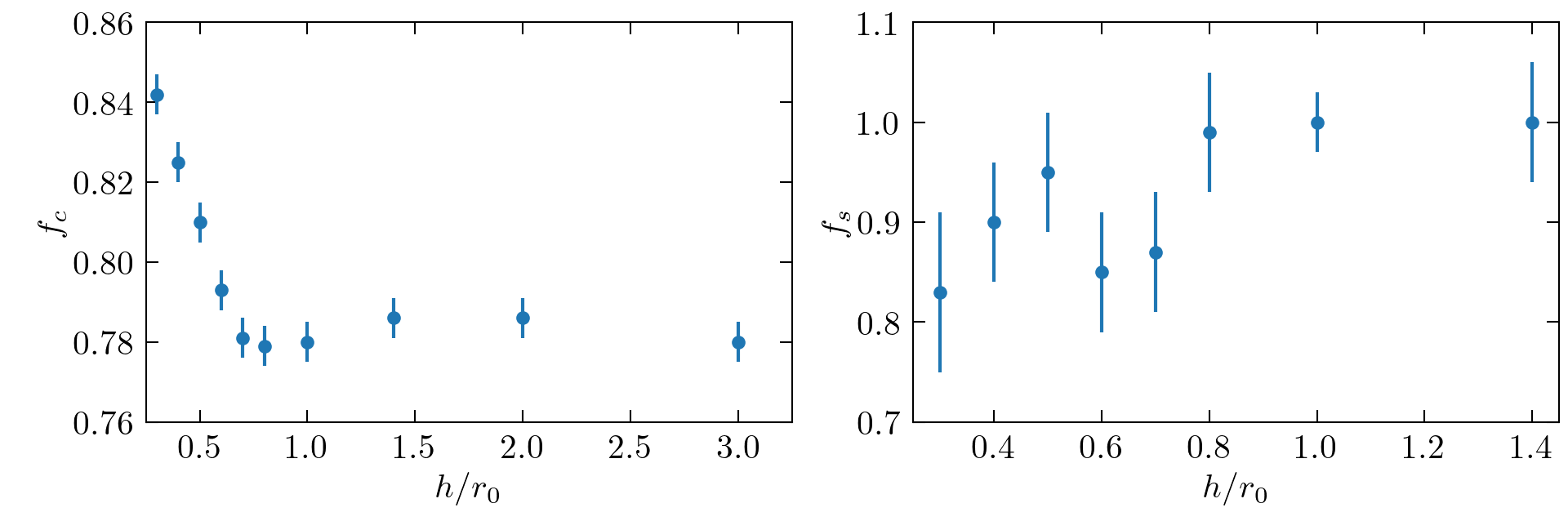}
\caption{Intra-layer condensate fraction and superfluid fraction of the 2D antidipolar bilayer liquid for different values of the inter-layer separation $h$ and $C_{12}/(E_0 r_0^{12}) = 0.103$, $n r_0^2 = 0.075$.}
\label{fig5}
\end{figure*}

Our main focus is to study the effect of the inter-layer interactions in the 
properties of 
the two-dimensional liquid, mainly, the equation of state (which allows us to 
obtain the equilibrium energy and density of the liquid), the superfluid 
fraction, and the condensate fraction. To do so, we calculate all these 
quantities for different values of the inter-layer separation $h$. We first 
report the results for the equation of state of the bilayer liquid, which we 
show in Fig.~\ref{fig4}. From these results, we see that the equation of state 
has a non-monotonous behaviour with respect to the inter-layer separation. As 
$h$ decreases, the EOS first shifts downwards, yielding lower energies per 
particle while leaving the equilibrium density mostly uneffected. Then, for 
values $h \lesssim 0.75 r_0$ this trend changes, with the EOS shifting to larger 
energies and the equilibrium density decreasing up to $n_{\rm eq} r_0^2 \simeq 
0.025$ for $h = 0.3 r_0$, which represents a reduction in a factor of 3 with 
respect to the single layer value. 

That non-monotonous behavior can be 
understood by evaluating the different contributions to the energy per particle 
of the system, which can be written as $E/N = E_K/N + U_{\rm intra}/N + U_{\rm 
inter}/N$, where $E_K$ is the total kinetic energy, $U_{\rm intra}$ is the 
potential energy of intra-layer interactions and $U_{\rm inter} = \frac{1}{2} 
\bra{\Psi} \sum_{i=1}^{N_1} \sum_{j=1}^{N_2} V^{\rm 12}({\bf r}_i - {\bf r}_j) 
\ket{\Psi}$. The different contributions to the energy are shown in 
Fig.~\ref{fig4b} for the case with $n r_0^2 = 0.075$. From the figure, we see 
that the kinetic energy term and the potential energy (mainly, the inter-layer 
potential energy) follow an opposite trend with respect to $h$: while the 
kinetic energy increases with $h$, the potential energy, which is negative, 
decreases. As the inter-layer distance decreases, the decrease of the 
inter-layer energy outweights the increase in kinetic energy, leading to a 
decrease of the total energy per particle. However, for $h \leq 0.8 r_0$ the 
increase in kinetic energy takes over, which leads to an increase in $E/N$. 
Looking back at Fig.~\ref{fig4}, we can also state that there are no significant 
qualitative changes in the EOS for $h > 1.4 r_0$ for the interaction considered.

Alongside the equation of state of the liquid, it is interesting to characterize 
the 
effect of the inter-layer interaction on the superfluid and 
Bose-Einstein condensation properties of the system, as these are of general 
interest in the field of ultracold atoms. We show in Fig.~\ref{fig5} the 
superfluid fraction of a layer along the $x$-$y$ plane for a 2D density of $n 
r_0^2 = 0.075$, $C_{12}/(E_0 r_0^{12}) = 0.103$, and different values of $h$. 
We fix the density to the value $n r_0^2 = 0.075$. From our DMC results, we see 
that $f_s$ remains close to 1 for inter-layer separations $h \gtrsim 0.8 r_0$, 
while for $h \lesssim 0.8 r_0$ it starts to decrease.
From the perspective of a given layer, the other
one exerts an external drag force that reduces superfluidity. We also report in 
Fig.~\ref{fig5} the condensate fraction of a single layer, $f_c$, as a function 
of the inter-layer distance.  We observe a small increase
($\lesssim 10\%$) of $f_c$ as $h$ decreases from $h \lesssim 0.8 r_0$. If one 
looks at Fig.~\ref{fig4} one can see that this value of the inter-layer distance 
is exactly the threshold at which the energy per particle changes its variation 
trend with respect to $h$, approaching lower energies per particle in absolute 
value. This may imply that the system is effectively closer to an ideal gas, which
makes the condensate fraction increase.

Let us put into an experimental context the inter-layer separations where 
inter-layer 
effects manifest. For our calculations to be valid, we need to consider an 
experimental system such that $h \gg l$, with $l$ the harmonic oscillator length 
along the $z$-axis, such that we can neglect exchange effects between layers. As 
an example, in the experiment of Ref.~\cite{lidu24:science} we have $l \simeq 2 
r_0$ for $^{164}$Dy atoms.
Since our calculations involve $h \sim r_0$, we would need an experimental
set-up such that $l \ll r_0$ to observe the effects reported in Figs.~\ref{fig4} 
and~\ref{fig5}, which is out of reach for current state of the art experiments. 
In particular, for the set-up of such experiment~\cite{lidu24:science}, our 
results would match the case with $h \gg 2 r_0$, which show no noticeable 
inter-layer effects in the equilibrium properties.


\section{\label{sec:conclusions}Conclusions}

In conclusion, we have studied the ground state properties of a two-dimensional 
system of 
particles under an \textit{antidipolar} interaction, i.e., a dipolar 
interaction with the reversed sign. We have considered a single 2D layer as 
well as a bilayer configuration. For the single layer and different interaction 
strengths, we have shown that the system behaves as a dilute liquid at 
low densities, with the equilibrium density being sensitive to the 
inter-particle interaction, which opens the way to tune it via a Feshbach 
resonance. At high densities, the system transitions to a solid phase with one 
atom per site  via a first order phase transition. 
Remarkably, this happens at a 2D density significantly lower than in a regular 
2D dipolar system. In the study of the bilayer geometry, we have restricted 
ourselves to the liquid phase, and have observed an inter-layer influence in the 
energy per particle, condensate fraction, and superfluidity of the system for 
sufficiently low inter-layer distances.

Our work sheds light into the physics of systems with antidipolar interactions, 
which are 
experimentally feasible either by rapidly rotating magnetic dipoles or by tuning 
the dipolar interactions of the recently realized polar molecules through 
microwave fields. Future perspectives include the investigation of inter-layer 
effects in the crystallization transition and the study of specific polar 
molecule interaction potentials, which have been recently proposed. This is of 
special relevance since polar molecules can achieve extremely large antidipolar 
strengths that cause the system to self-organize in two-dimensional 
structures~\cite{zhang26:nature,langen25:prl}, enabling the potential study of 
the interplay of strong interactions and physical phenomena exclusive to the 
two-dimensional regime without the need of any external strong trapping.

\section{\label{sec:acknowledgements}Acknowledgements}

We acknowledge support by the Spanish Ministerio de Ciencia, Innovación y Universidades (grant PID2023-147469NB-C21, financed by MICIU/AEI/10.13039/501100011033 and FEDER-EU).

\bibliography{paper_2d_antidipoles}

\end{document}